# Investigation of re-entrant relaxor behaviour in lead cobalt niobate ceramic

Adityanarayan H. Pandey*, Surya Mohan Gupta*

*Homi Bhabha National Institute, Anushaktinagar, Mumbai-400094, India.*
*Laser Materials Section, Raja Ramanna Centre for Advanced Technology, Indore-452013, India.*
{*Email: anbp.phy@gmail.com(A. H. Pandey); surya@rrcat.gov.in(S. M. Gupta)}

**Abstract:** The temperature dependent dielectric properties revealed re-entrant relaxor behaviour ($T_m$ ~130 K and 210 K for 1 kHz) below a high temperature diffused phase transition, $T_c$ ~270 K in lead cobalt niobate (PCN). Multiple positive/negative magnetodielectric effect and deviation from straight line at ~130 K is observed in temperature dependence of inverse susceptibility, which depicts origin of frustration. Microstructure examination depicts closely packed grains with grain size ~8-10 μm and XRD pattern revealed single phase pseudo cubic crystal structure having Pm3m symmetry with lattice constant ~4.0496(2) Å. Rietveld Refinement on XRD data yields larger value of thermal parameters, implying Pb and O are disordered along <111> and <110> directions respectively. Observation of $A_{1g}$ (780 cm$^{-1}$) mode in Raman spectroscopy and F-spot in SAED pattern along <110> unit axis in TEM suggests presence of nano scale 1:1 Co and Nb non-stoichiometric chemical ordering (CORs), akin to lead magnesium niobate (PMN). K-edge XANES spectra reveals the presence of cobalt in two oxidation states ($Co^{2+}$ and $Co^{3+}$); whereas, niobium exists in $Nb^{3+}$ state. Therefore, these local-average structural properties suggest chemical, structural and spatial heterogeneities. Such multiple heterogeneities are believed to play a crucial role in producing re-entrant relaxor behaviour.

## 1. Introduction

Relaxor ferroelectrics (RFEs) are important multifunctional materials that have received enormous attention in recent years due to their extraordinary properties viz. large dielectric constant with large frequency dispersion, hysteresis free polarization, electrostrictive coefficient and electro-optic properties, which are useful for multilayer capacitors, transducers, actuators, sensors, micropositioners, motors, and light valves etc. [1,2] Family of Pb-based mixed perovskite relaxors with general formula Pb(B'$_x$B"$_{1-x}$)O$_3$ where B' is lower valence cation (e.g. $Mg^{2+}$, $Co^{2+}/Co^{3+}$, $Ni^{2+}$, $Zn^{2+}$, $Sc^{3+}$, $Fe^{3+}$, $In^{3+}$ etc.) and B" are higher valence cations and ferroelectrically active ions (e.g. $Ti^{4+}$, $Zr^{4+}$, $Nb^{5+}$, $Ta^{5+}$, $V^{5+}$, $W^{6+}$ etc.) are widely studied. Frequency dependent diffuse dielectric maximum and the temperature of dielectric maximum ($T_m$) are typical dielectric characteristics of the relaxors. The $T_m$ does not correspond to any phase transition from paraelectric to long range ordered ferroelectric state. Number of models are reported to explain some of the relaxor dielectric characteristics but none has explained all the dielectric and ferroelectric properties of the relaxor [3,4]. Lead magnesium niobate (PMN) is widely studied Pb-based mixed relaxor ferroelectric and well known for very high dielectric constant and least hysteresis loss, which is important for actuators, capacitors and deformable mirror application. At present it is clear that broad dielectric response and hysteresis free polarization behaviour of relaxors is correlated with statistical distribution of polar nano regions (PNRs) having symmetry mostly R3m distributed in the paraelectric matrix having Pm3m crystal symmetry [5-7]. These PNRs initially forms at burns temperature ($T_b$), which is much higher than the $T_m$. To distinguish the PNR state below $T_b$ from the paraelectric state above $T_b$, it is termed as ergodic relaxor (ER) state. The number and size of the PNRs are reported to increase with on cooling below the $T_b$. Selected area electron diffraction along <110> unit axis and transmission electron dark field image has revealed faint superlattice reflections at ½<111> and nano-meter sized white regions, which are related to the B-site cation chemical ordering and these chemical ordered regions (CORs) are reported to present along with the PNRs. These CORs are believed to be the source of quenched random field, which does not allow PNRs to grow into long range ordered regions below $T_c$ [9]. This state of relaxors below $T_c$ is called nonergodic relaxors (NR) state due to its aging, an anomalously wide relaxation time spectrum, and thermal and history dependence characteristics of nonergodic behaviour [7-9]. A long range ferroelectric state in these relaxors is induced below $T_c$ with the application of high electric field or mechanical strain or thermal annealing or even by chemical substitution [10-13].

Out of the many relaxors, three Pb-based disordered relaxor niobates [$Pb(B'_xNb_{1-x})O_3$], where B' site is occupied by magnetic transition metal Fe, Co and Ni magnetic ions are potential candidate to show multiferroic properties. Lead iron niobate [$Pb(Fe_{1/3}Nb_{2/3})O_3$] is extensively studied but lead cobalt(II) niobate [$Pb(Co_{1/3}Nb_{2/3})O_3$, PCN] is one of the oldest but least studied members of this group. First report on synthesis and dielectric characterization of PCN single crystal has reported cubic crystal structure with centre position of $O_6$ octahedral is randomly occupied by $Co^{2+}$ and $Nb^{5+}$ ions at room temperature [14]. Dielectric study has revealed relaxor behaviour with strong frequency dispersion, $\varepsilon'_m$ ~6000 at $T_m$ ~-70 °C (for 1 kHz) and temperature dependent PE loop indicating polarization switching due to domain reorientation in sufficiently large field [14]. There are number of contradiction reports on the dielectric maxima and temperature of dielectric maxima for single crystal and ceramic of PCN [14-20]. Large deviation in $T_m$ for single crystal and ceramic is still not clear. Solid solution of PCN with PZT and PT are also studied to enhance ferroelectric as well as piezoelectric properties [21,22]. There are contradictory reports on temperature dependent magnetization studies of the PCN, e.g., Popova *et al*. has reported antiferromagnetic transition temperature $T_N \sim 130$ K but only paramagnetic behaviour is reported for single crystal and ceramic down to 2 K without any short range ordering [17,19].

The purpose of this work is to synthesise PCN and characterize its compositional, microstructural, electric and magnetic properties and establish correlations between the dielectric and magnetic ordered state. It is already known that the Co-ion can exist in multiple valance as well as spin states, which may give interesting magnetic and dielectric properties. Comprehensive room temperature structural analysis of PCN using synchrotron x-ray diffraction (XRD), x-ray absorption spectroscopy (XAS), and RAMAN spectroscopy and its correlation with dielectric spectroscopy, ferroelectric and magnetic properties reveals a re-entrant phenomenon in PCN, which is explained by breaking of ferroelectric order by development of magnetic correlations below 150 K.

## 2. Experimental details

Lead Cobalt Niobate (PCN) is synthesized by the Columbite precursor method using high purity PbO (99.9%), CoO (99.99%), and $Nb_2O_5$ (99.9%) [23,24]. The Columbite precursor, cobalt niobate ($CoNb_2O_6$), is prepared by mixing the stoichiometric amounts of CoO and $Nb_2O_5$ along with reagent grade ethanol in planetary ball mill (P6 Fritsch) for 18 hours using zirconia grinding media. The slurry is dried in oven at 80 $^o$C and the dried powder is then calcined at 1200 $^o$C in air for two hours. The calcined powder is ball milled again for 18 hours to achieve homogenous PCN powder. Single phase of $CoNb_2O_6$ is confirmed by indexing X-Ray Diffraction (XRD) peak with JCPDS file (72-0482). Further, predetermined amount of $CoNb_2O_6$ powder and PbO (~2% excess for compensating lead loss during calcination) is ball milled, dried and calcined at 900 $^o$C for 4 hours in closed alumina crucible. The calcined powder is mixed with binder polyvinyl alcohol (PVA-3mol%) before pressing into pellets (15 mm diameter with 4-5 mm thickness) under a uni-axial hydrostatic pressure of 200 MPa. These pellets are sintered at 1200 $^o$C for 2 hours after burning out of the binder at 450 °C for 4 hours in closed alumina crucible. The sintered pellet has a sinter density greater than 98% of the theoretical value.

The calcined and sintered powder samples are analyzed by using Bruker powder diffractometer (Cu K$_\alpha$ source with λ = 1.54 Å) at a scan rate of 0.5 °/min with 0.01° step size. A small piece of the sintered sample is grounded well and annealed at 600 °C for 12 h to remove the grinding induced strain before taking the XRD patterns. The Fullprof software is used to refine the lattice parameters with the XRD data [25]. The sintered block is then cut into slim disks and polished using emery papers to acquire smooth parallel surfaces. The smooth surfaces are ultrasonically cleaned to eliminate the dust particles. Thin pellet is then electrode using gold sputtering, succeeded by application of silver paste (fired at 450 °C for 2 min). Dielectric properties are recorded using a 6505B precision impedance analyzer (Wayne-kerr instrument, which can cover a frequency range of 20 Hz - 5 MHz). For low temperature measurements, the sample is placed in a cold finger set-up, which can be operated between 80 K and 450 K. The temperature is determined using a temperature controller with a RTD mounted on the ground electrode of the sample holder. The test chamber, analyzer and Eurotherm temperature controller are interfaced with a computer to record data at 50 different frequencies while heating at a rate of 2 K/min. P-E hysteresis loop is measured at 50 Hz using Precision workstation of Radiant Technology, USA.

Grain size, morphology of fractured surface is imaged using field emission ccanning electron microscope (FE-SEM, Carl Zeiss, SIGMA) equipped with energy dispersive spectroscopy (Oxford Inca X-Act LN2 free). Gold is sputtered on the fractured surfaces of ceramic sample. TEM samples are prepared by ultrasonically drilling 3-mm discs which were mechanically polished to ~100 μm. The center portions of these discs are then further ground by a dimpler to ~10 μm, and argon ion-milled (operating at 2-6 KeV) to perforation. The TEM studies are done on a Phillips CM200 microscope operating at an accelerating voltage of 200 kV. Room temperature Raman spectra of the sintered sample is recorded by using LABRAM HR-800 spectrometer equipped with a 488 nm excitation source and a CCD detector giving real spectral resolution of better than 0.5 cm$^{-1}$. The Jandel 'peakfit' software is used to deconvolute the overlapping modes. The fitting of the Raman spectra is accomplished by using Pseudo-Voigt peak functions (PV = $p$*L + (1-$p$)*G, 0≤ $p$ ≤1, where L and G stand for Lorentzian and Gaussian, respectively) to determine the characteristic parameters of all the Raman peaks, like peak position, full width at half maximum (FWHM) and intensity. Field dependent magnetic measurements are carried out using MPMS SQUID (make Quantum Design, USA) magnetometer. Room temperature X-ray absorption near edge structure (XANES) is recorded in fluorescence mode (for Co and Nb K-edges) at Scanning EXAFS Beamline (BL-9) at the Indus-2, India. The photon energy is calibrated by the Co K-edge XANES spectra of standard Co metal at 7709 eV for Co edge and Nb K-edge XANES spectra of standard Nb metal at 19000 eV. The fluorescence XANES spectra are recorded using vortex energy dispersive detector (VORTEX-EX).

## 3. Results and discussion

### 3.1. Microstructural and structural studies

Microstructure observed under scanning electron microscope (SEM) of fractured surface of PCN ceramic is shown in Fig. 1. Inter granular fractured grains 10-15 micro meter in size are clearly visible. No different grain morphology corresponding to the secondary phase or large porosity is noticed. Entrapment of small spherical pores within the grain is observed, which is marked by an arrow, agrees well with the density measurement by the Archimedes liquid displacement method and assured the sintered density of the PCN ceramic. X-ray diffraction pattern of sintered PCN ceramic is shown in Fig. 2. The detailed structural information is revealed with the Rietveld refinement of room temperature XRD pattern of PCN. Pseudo-Voigt function is used to define peak shape. The background is modelled using a fifth order polynomial and scale factor, zero correction, background, lattice parameter, half-width, position co-ordinates, isothermal, asymmetry, composition parameters are refined. All the diffracted peaks of PCN are indexed by considering pseudo cubic Pm-3m symmetry for structural refinement. Ideal position of atoms in the perovskite, i.e. $Pb^{2+}$ occupies 1(a) site at (0,0,0) positions, $Co^{2+}/Nb^{5+}$ occupies 1(b) site at (1/2,1/2,1/2) positions and $O^{2-}$ occupies 3(c) site at (1/2,1/2,0) positions are considered. Figure 2 compares the XRD pattern of the PCN powder with the simulated X-ray diffraction pattern. The red colour points represent the experimental data and black line is the simulated data. The bottom blue line represents the difference between the experimental and simulated diffraction patterns revealing reasonable matching of the experimental data with the simulated profile of the XRD pattern. The quality of Rietveld refinement is adjudged based on the minimal value of agreements factors i.e. $R_p$, $R_{wp}$, $R_{exp}$, $\chi^2$, $R_b$, $R_f$ [25]. The corresponding agreement factors are tabulated in Table 1 confirming a reasonably good fit. The calculated lattice parameter, 'a' = 4.0496(2) Å is consistent with the earlier reported PCN single crystal and ceramic samples [17]. It may be noted from the Table 1 that the thermal parameters of $Pb^{2+}$ and $O^{2-}$ are reasonably higher than that of $Co^{2+}/Nb^{5+}$. This indicates that $Pb^{2+}$ and $O^{2-}$ may be highly disordered, similar to other already reported Pb based mixed perovskite PMN and PFN [17,25]. Popova *et al.* [17] have also observed large value of thermal parameter for $Pb^{2+}$ and $O^{2-}$ ions in single crystal PCN, which has been related to structural disorder in the form of displacement. To improve upon the thermal parameters of $Pb^{2+}$ and $O^{2-}$ ions, a split-atom approach is used. In this approach, $Pb^{2+}$ and $O^{2-}$ ions are allowed to shift statistically in various crystallographic directions rather than fixing them on high symmetry cubic positions. Significant improvement in the thermal parameters have been observed when atomic positions of Pb and O are isotropically shifted along <111> and <110> directions, respectively. The isotropic shift of ~ 0.045 Å for $Pb^{2+}$ along <111> direction and ~ 0.046 Å for $O^{2-}$ along <110> direction has improved isothermal parameters by an order of magnitude. The split-atom approach has also been reported to improve the thermal parameter in Rietveld refinement of powder X-ray diffraction and neutron diffraction data of PMN [25]. Further Raman spectra analyzed on the basis of structural observation.

### 3.2. Raman spectroscopy

Raman scattering spectroscopy is an effective tool to study the local structures of Pb based mixed perovskite complex materials, because local symmetry of nano regions is different from that global symmetry and these are governed by different selection rules [26]. The B-site disorder, presence of the CORs and off-center B-site ion displacements results in multiple in-homogeneities causing difficulties in interpretation of Raman bands. Hence, origin of some of the Raman bands is still question of debate. Figure 3(a) represents room temperature Raman spectrum of PCN and for sake of clarity it is compared with well-known mixed perovskite compound PMN. For an ideal cubic perovskite $ABO_3$ structure with Pm3m space group, the first-order Raman modes are forbidden by symmetry. But, Fig. 3(a) shows many broad and overlapping Raman bands, which is consistent with presence of large disorder in PCN system similarly to that in the PMN [26-29]. Qualitatively both Raman spectra are similar at first glance, which is obvious because of the same structure. On comparing with the PMN spectrum, positions of few modes are found red shifted with larger intensity and FWHM. The Raman spectrum of the PMN is due to i) 1:1 non stoichiometric ordering of Mg/Nb along <111> direction with Fm3m space group in the CORs and ii) the PNRs having rhombohedral symmetry. For convenience, the Raman spectra of the PCN is divided into three regions [26], i) low frequency region (below ~ 150 cm$^{-1}$), where the bands are allotted to Pb-BO$_6$ stretching modes ii) medial frequency region (from ~150 to 500 cm$^{-1}$), where bands are due to combined B-O-B bending and O-B-O stretching modes, and iii) high frequency region (from ~500 to 800 cm$^{-1}$), where bands are assigned to B-O-B stretching modes. The strongest mode at ~780 cm$^{-1}$ is assigned to 1:1 Nb-O-Mg stretching mode corresponding to Fm3m and documented as A$_{1g}$, which is archetypical for presence of the CORs. Jiang *et al.* [30] have demonstrated that the A$_{1g}$ mode is very sensitive to B-site ordering and any change in ordering of B-site is reflected in this mode. The presence of this ordering mode is observed in many complex Pb-based mixed perovskites [30]. In contrast to this Woo *et al.* [31] has not observed any superlattice reflections at 1/2 <111> in the SAED pattern along <110> unit axis and the A$_{1g}$ mode in Raman spectrum for PbFe$_{1/2}$Nb$_{1/2}$O$_3$. Figure 3(a) shows the presence of A$_{1g}$ mode ~ 780 cm$^{-1}$ for the PCN and this mode is observed red shifted and broadened when compared to A$_{1g}$ mode of the PMN. The red shift of the A$_{1g}$ and other modes can be explained using simple harmonic approximation model, ω = (k/m*)$^{1/2}$ where ω is Raman shift, k is bond constant and m* is effective mass of modes. Due to similar lattice parameter of the PCN and PMN, i.e., 4.049 Å and 4.047 Å, respectively, the Raman shift will relate inversely with the effective mass. Since the atomic mass of Co is higher than that of Mg; hence all the modes involving Co are found red shifted.

The origin of Raman bands group theory analysis is carried out on the basis of the structural result and available structural literature of Pb-based mixed perovskites. There are three possible symmetries present in PCN, namely, i) globally, Pm3m symmetry supported with the XRD result ii) locally, R3m symmetry supported with the disorderness in the position of Pb and O-ions, imply the presence polar nano regions (PNRs) and iii) locally, Fm3m symmetry with the presence of CORs supported with the A$_{1g}$ mode. Group theory predicts the presence of 20 modes for PCN, 4 modes originating from Fm3m symmetry and 16 modes originating from R3m symmetry [28,29]. The room temperature Raman spectrum is de-convoluted using pseudo-voigt distribution peak profile. Figure 3(b,c) shows the de-convoluted PCN Raman spectra in 50-400 cm$^{-1}$ and 400 to 800 cm$^{-1}$ wavength range for clarity. It may be noted that few soft Raman modes around ~50 cm$^{-1}$ are not visible due to experimental limitation. The difference between observed and theoretically predicted modes is due to overlapping of Raman modes and presence of disorder, which makes it impossible to distinguish all modes. These results are consistent with the Raman study reported for PMN and PMN-PT ceramics [28,29]. Therefore, the Raman study reveals the presence of nano scale 1:1 non stoichiometric order of B-site cations regions (CORs) and the rhombohedral symmetry polar nano regions (PNRs) distributed randomly in the matrix of cubic symmetry.

### 3.3. TEM

Presence of the CORs and the PNRs in PCN ceramic is directly visualized using <110> zone selected area electron diffraction (SAED) pattern and with bright field TEM imaging, respectively [32]. Figure 4(a,b) represents the bright-field and SAED pattern imaging of PCN ceramics. Local random contrasts are seen in Fig. 4(a), which represents the polar nano-domains at room temperature. The average size of these polar nano-domains is less than 5 nm and consistent with its presence below the Burns temperature [32-35]. Figure 4(b) displays the <110> zone SAED patterns of PCN. Along with the allowed strong reflections, which are originating from the basic perovskite structure, extra weak spots at (½½½) reciprocal positions along <111> (F-spots, super-lattice reflections) are also apparent, as marked by arrow. These super-lattice reflections indicate local chemical-ordering in some regions of the PCN sample, which causes doubling of the unit cell. Stoichiometric 1:2 ordering of Co/Nb is expected as per the chemical formula of PCN ($PbCo_{1/3}Nb_{2/3}O_3$) which has been reported for $BaMg_{1/3}Nb_{2/3}O_3$. Instead of 1:2 stoichiometric ordering, presence of the weak superlattice reflection is related to non-stoichiometric 1:1 ordering, similar to that has been reported for the PMN [32-35]. It is believed that Co and Nb are present in alternate layers stacked along <111> as reported by the '*random layer model*' [34,35]. Non-stoichiometric ordering of the PMN should be different than the ordering of the PCN because two different oxidation states of the Co-ion is known to be quite stable. In order to determine the oxidation state of Co-ion at B'-site, x-ray absorption near edge spectroscopy (XANES) is carried out.

### 3.4. XANES

XANES is an element specific spectroscopic tool which provides information about oxidation states, local coordination and electronic structure (hybridization effect of orbitals) of the elements present in the sample [36]. Edge step normalized Co K-edge XANES spectra of PCN is compared with the Co metal foil, cobalt oxide and CoF3 standards in Fig. 5(a). K-edge XANES spectrum of Co-ion relates the transitions, 1s→4p and 1s→3d to main and pre-absorption edge, respectively. The pre edge feature of XANES spectrum shows crystal field splitting of 1.0 (± 0.04) eV between $e_g$ and $t_{2g}$ states, indicating a mixture of high spin states of both $Co^{2+}$ and $Co^{3+}$ and is shown in upper inset of the Fig. 5(a) [37-39]. Energy position of the main absorption line is determined from the maximum energy value of first order differentiated spectrum, which provides the oxidation state of the absorbing atom. It is clear from Fig. 5(a) that the main edge corresponding to PCN lies between that of standard CoO and $CoF_3$ samples, which indicates that Co-ion in the PCN sample is not entirely in $Co^{2+}$ state. A simple linear combination formula is generally used to calculate the concentration of $Co^{2+}$ and $Co^{3+}$,

Energy positions of PCN sample = {Energy position of CoO × $x$ + Energy position of $CoF_3$ × (1-$x$)} /100"  Eq. (1)

Here, $x$ is the calculated concentration of $Co^{2+}$. This formula assumes a linear dependence of the chemical shift on the average valence and the edge energy positions of PCN, CoO and $CoF_3$ are plotted in the inset of lower part of the Fig. 5(a). The relative concentrations of $Co^{2+}$ and $Co^{3+}$ is calculated from Eq.1 as ~ 38% and 62%, respectively.

Figure 5(b) shows step normalized XANES spectra for Nb K-edge 1s→5p with two standard references for $Nb^{4+}$ ($NbO_2$) and $Nb^{5+}$ ($Nb_2O_5$). The XANES spectra reveals Nb existing in multiple oxidation state $Nb^{3+}/Nb^{4+}/Nb^{5+}$. The XANES measurement of Columbite precursor, cobalt niobate ($CoNb_2O_6$) at Co and Nb K-edges reveals that Co exists in multiple valence state ($Co^{2+}/Co^{3+}$) and Nb exists in $Nb^{5+}$ oxidation state in the Columbite precursor.

### 3.5. Dielectric spectroscopy

Figure 6(a-c) illustrates the temperature variation of real [$\varepsilon'(T)$], imaginary [$\varepsilon''(T)$] parts of dielectric permittivity and loss tangent [$\tan\delta(T)$], measured in temperature range of 80 K to 480 K and at different frequencies in the range of 100 Hz to 1 MHz. Dielectric and loss spectra of the PCN show three distinctive dielectric anomalies marked I, II, III and at higher temperatures (above region III) dielectric loss suddenly rises due to conduction losses. The dielectric behaviour of PCN matches well with earlier reports on PCN ceramic [17,20]. As the sample is cooled the dielectric constant increases and a broad diffused maximum in dielectric constant ($\varepsilon_m$ ~4400) is observed at temperature $T_m$ ~260 K, which indicates the existence of a phase transition [15]. Below $T_m$ ~260 K, there are two $\varepsilon''$ -$T$ peaks marked by I (~120 K) and II (200 K), which are clearly visible in Fig. 6(b). The $\varepsilon''_{m1}$ and $T''_{m1}$ increases with increase of frequency similar to relaxors. In region II, $\varepsilon''_{m2}$ and $T''_{m2}$ behaves akin to region I but this hump is present only above 10 kHz frequencies. On further cooling below $T_m$, dielectric constant decreases down to 80 K with an anomaly near 160K and below this frequency dependent $\varepsilon'$ is observed. The observed low temperature glassy phase appeared below phase transition temperature (~250 K) is called re-entrant phase [40].

Figure 6(d) shows $1/\varepsilon$-$T$ plot shows deviation from its linear variation near ~320 K. Therefore, the temperature variation of dielectric constant of PCN above 320 K is fitted by Curie-Weiss law (Eq. 2) and broad diffused maximum is described using quadratic form of Curie-Weiss law (Eq. 3) given as follows [41]

$$1/\varepsilon(T, \omega) = (T-T_{cw})/C \qquad (2)$$

$$\varepsilon_A(\omega) / \varepsilon(T, \omega) = 1 + (T - T_A(\omega))^2 / 2\delta_A^2 \qquad (3)$$

where, $C$ is the Curie constant, $T_{cw}$ is Curie-Weiss transition temperature, $\varepsilon_A$ ($> \varepsilon_m$), $T_A$ ($< T_m$) and $\delta_A$ are fitting parameters, practically independent of frequency and valid for long range of temperatures [41]. Figure 6(d) shows the fitted curve of $1/\varepsilon$ vs T by Eq. 2 and 3; yield parameters, $C = 2.6 \times 10^5$ K, $T_{cw} = 250$ K, and $\delta_A \sim 90$. The value of $T_{cw}$ is close to the temperature of the $\varepsilon_m$ 260 K implying paraelectric to ferroelectric phase transition temperature. It is noticed that the value of degree of diffuseness, $\delta_A$ is almost double of well-known relaxor PMN ($\delta_A \sim 45$) implying larger disorder at the B-site in PCN ceramic [32]. It has already been revealed that an additional disorder is induced when Gd-substitutes the Mg-site, resulting in higher value of degree of diffuseness ($\delta_A$). It is believed that the presence of different oxidation states of Co and Nb-ions at B-site is resulting high degree of diffuseness.

It is noted that the effect of frequency dependent relaxation in all three regions is prominent in imaginary part of permittivity and not clearly observed in real part of dielectric permittivity, which may be due to masking of the real feature by high conduction losses. Figure 5(e) depicts the temperature dependence of relaxation frequency, $\omega = \omega(T"_m)$ and it's fitting with cluster glass model [42]. The cluster glass model (Eq. 4) is reported, based on critical slowing down dynamics of PNRs, which is well known in magnetic cluster glasses and structural glasses.

$$\omega = \omega_o (T_m/T_g - 1)^{zv} \qquad (4)$$

where $\omega_o$ is the Debye frequency ($\tau_o = \omega_o^{-1}$ is the microscopic time associated with flipping of fluctuating dipole entities), $T_g$ is glass transition temperature also called blocking temperature and $zv$ is critical dynamic exponent for the correlation length. A good agreement between fitted curve (solid line) and the experimental data (open circle) can be observed for two I and II regions. The parameters presented in Table 2 are well within the limit of its physical significance.

A sudden increase in the tan δ is observed in region III and is shown in Fig. 5(c), which is associated with hopping electrons or/and ionic conduction loss. To determine the different conduction processes overlapping in regions III, temperature dependence of ac conductivity is calculated using Eq. 5.

$$\sigma_{ac}(T) = \omega_o \varepsilon_o \varepsilon'(T) \tan\delta(T) \quad (5)$$

The plot of ln $\sigma_{ac}$ vs 1000/$T$ clearly shows different slopes below and above 250K and is shown in Fig 6(e). The data points are related with the Arrhenius relation and activation energies corresponding to various thermally activated processes is calculated. The activation energies in different temperature regimes are $E_{a1}$ ~0.471(4) eV and $E_{a2}$ ~0.196(3) eV. The activation energies in the range 0.1-0.3 eV is reported for localized hopping of polarons and 0.3-0.5 for and 0.6-1.2 eV is associated to single-ionized and doubly-ionized oxygen vacancies, respectively [43]. Similarly, the activation energy, $E_{a1}$ ~0.471(4) eV is associated with hopping of single-ionized oxygen vacancies and $E_{a2}$ ~0.196(3) eV with the two-site polaron hopping process of charge transfer between $Co^{2+}$-$Co^{3+}$ sites [44].

### 3.6. Polarization

Figure 7(a-c) shows P-E loop traced for PCN at 275, 180 and 80 K when 10 kV/cm external field is switched at 50 Hz. The *P-E* loop at 275 K is consistent with switching of field induced polar regions displaying low coherent length among the polar regions. On cooling to 180 K, the coherent length increases and the cooperative interaction leads to anti-ferroelectric phase transition, which is clearly shown by double *P-E* hysteresis loop at 180 K. Further cooling to 80 K, leads to development of frustration among the polar regions which resulting into critically slowing down of the polar nano-regions dynamics. Figure 7(d) compares the temperature dependence of $P_{max}$, $P_r$ and $E_c$ revealing the paraelectric to antiferroelectric transition and then the coupling between the polar regions become frustrated below 150 K leading to reduction of polar region size, i.e., re-entrant behaviour. The dynamics of the polar nano-regions slows down below 110 K, which is consistent with reduced $P_{max}$ and hysteresis loss. The frustration is believed to develop near 150 K, where anti-ferromagnetic coupling is reported.

### 3.7. Magnetization

Temperature dependence of zero-field-cooled (ZFC) dc susceptibility measurement in temperature range of ~5 K to 300 K under an applied field of ~ 100 Oe is carried out. There are contradictory reports about the magnetic ordering of PCN, e.g., Venevtsev *et al.* has reported AFM phase transition at $T_N$ ~130 K but Chillal *et al.* have observed paramagnetic behaviour in single crystal of PCN with an additional anomaly near 50 K, which is reported to be suppressed by the application of high magnetic field [18,19].

Figure 8(a) shows temperature dependent susceptibility [$\chi(T) = M/H$] and inverse susceptibility ($1/\chi$) for PCN ceramic. Any anomalies in $\chi(T)$ near ~130 K or 50 K is not observed between 5-300 K. However, a deviation from the Curie-Weiss fitting is noted at ~150 K in the $1/\chi$ (T) plot. The deviation from linearity is due to development of weak correlation between the magnetic moments of $Co^{2+}/Co^{3+}$, which may not be strong enough to grow to long range ordered phase upon cooling.

The Curie-Weiss law is fitted in linear region of $\chi^{-1}(T)$ i.e. in the temperature range of ~170 K to 300 K and the Curie–Weiss temperature ($\theta_p$), the Curie-Weiss constant (*C*) and effective magnetic moment ($\mu_{eff}$) are calculated. The Curie-Weiss constant and effective magnetic moment 1.172 emu-K/mol and 5.33 $\mu_B$ are calculated, which is consistent with earlier report [37-39]. The large negative value of -84.1 K for $\theta_p$ reveals local weak predominant AFM interactions in the PCN. These results are consistent with the earlier reported results in which no long range magnetic ordering is observed. It is not clear why the magnetic transition from paramagnetic to AFM is not observed because the concentration of Co-ions is well above the percolation threshold required to form long range magnetic order in the disordered PCN. Here, the percolation threshold means the minimal concentration of magnetic ions distributed in non-magnetic matrix below which the individual magnetic moments do not correlate to form long range order. The absence of long range magnetic ordering is believed to be due to the presence of enhanced degree of disorder at the B-site. The magnetization versus magnetic field (*M-H*) curve recorded at 300 K and 5 K are compared in Fig. 8(b). The room temperature M-H curve of the PCN shows linear dependence between magnetization and magnetic field and non-linear dependence, which tends to saturate at low temperature 5K suggesting locally anti-ferromagnetic regions but globally paramagnetic nature [Fig. 8(b)]. The magnetic correlation is believed to develop at low temperature which may be originating from super-exchange correlations between the $Co^{2+}$-$Co^{3+}$ or $Co^{2+}$-$Co^{2+}$ and $Co^{3+}$-$Co^{3+}$ [19,37,38].

### 3.8. Magnetodielectric effect

Figure 9(a) compares the temperature dependent dielectric constant and loss (at 10 kHz frequency) of PCN ceramic in the present and absence of the 9 T magnetic field. The temperature dependent dielectric permittivity, $\varepsilon'(T)$ clearly reveals an abnormally near 150 K, which is related to the development of magnetic correlations leading to reduction in the correlation among the polar regions leading to re-entrant behaviour.

Figure 9(b) compares the temperature dependent magnetodielectric (%$MD$) and magneto-loss (%$ML$) coefficient for PCN ceramic sample. Below 300 K, the MD first increases up to ~4% near ~160 K and then decreases to -3% below 160K, which then increases to 2% around 40 K. The $ML$% remains negative in all the temperature except between 25 to 50 K. It is believed that weak interaction of the magnetic field with magnetic moments of moving nano-domains walls is responsible for small magneto-dielectric effect. The change from negative to positive MD effect around 250 K seems to be related with the PNRs size and initiation of the magnetic field influence on the dynamic characteristics of the nano-domains walls. As these nano-domains grow in size, the MD effect increases up to a temperature 160 K and then decreases due to development of magnetic correlation. It may be noticed that frustration created by anti-ferromagnetic correlations, which starts to develop around 150 K is believed to result into re-entrant behaviour in PCN. The positive and negative MD behaviour are also observed in disordered double perovskite $Pr_2CoMnO_6$ ceramic whereas single %$MD$ peak is reported in B-site ordered phase [45]. Similarly, Imamura *et al.* [46] observed positive and negative MD effect in A-site ordered oxide, $(BiMn_3)Mn_4O_{12}$. They suggested that the anomalous MD behaviour is due to occupation of magnetic ion $Mn^{3+}$ at A and B both sites are the important crystal chemical factor.

## 4. Conclusion

Temperature dependent dielectric and ferroelectric properties revealed re-entrant low temperature relaxor behaviour near $T_m$ ~120 K for 1 kHz along with a diffused transition, $T_c$ ~250 K. The local and average structural properties suggest chemical, structural and spatial heterogeneities. K-edge XANES spectra analysis has revealed two oxidation states $Co^{2+}$ and $Co^{3+}$ in 38:62 ratio and Nb in $Nb^{3+}$ oxidation state. Raman spectroscopy suggests strong disorderness and presence of nano scale 1:1 Co and Nb non stoichiometric chemical ordering (CORs), which is supported by $A_{1g}$ (780 cm$^{-1}$) mode and presence of superlattice reflections at <½½½> in <110> SAED pattern. The multiple heterogeneities viz., chemical, structural and spatial observed by local and structure characterization are believed to play a crucial role in producing re-entrant relaxor behaviour. Magnetization and magnetodielectric (MD) effect of PCN ceramics has revealed bipolar magnetic field dependence on the dielectric properties. Weak anti-ferromagnetic correlations are believed to develop around 150 K in PCN, which results in re-entrant behaviour in PCN.

## Acknowledgements

Authors are grateful to Dr. A. K. Karnal and Dr. V. S. Tiwari for their constant support and encouragement, Dr. S. K. Rai for XRD measurement, Dr. V. G. Sathe for Raman mesurement, Dr. N. P. Lalla for TEM measurement, Dr. V. R. Reddy for P-E loop measurement, Dr. A. M. Awasthi for magnetodielectric measurement, Prof. A. K. Nigam for Magntic measurements, and Dr. G. Singh, and Mrs. R. Selvamani for useful discussions. Mr. Pandey acknowledges Raja Ramanna Centre for Advanced Technology, Indore for Senior Research Fellowship.

**Tables and Figures**

**Table 1.** Structure parameters obtained after Rietveld refinement of lead cobalt niobate ceramic sample for Pm3m crystal symmetry

| a (Å) | Wyckoff positions | | | | Isothermal parameter (Å$^2$) | Occupancy | Agreement Factors |
|---|---|---|---|---|---|---|---|
| 4.0496(2) | Atoms | $x$ | $y$ | $z$ | | | |
| | Pb | 0 | 0 | 0 | 3.298 | 1 | $\chi^2$ = 1.21, |
| | Co/Nb | 0.5 | 0.5 | 0.5 | 0.308 | 0.33/0.67 | R$_B$ = 3.89, |
| | O | 0.5 | 0.5 | 0 | 2.114 | 3 | R$_F$ = 4.52 |
| 4.0496(2) | Pb | 0.04458 | 0.04458 | 0.04458 | 0.210 | 1 | $\chi^2$ = 1.20, |
| Pb<111> | Co/Nb | 0.5 | 0.5 | 0.5 | 0.280 | 0.33/0.67 | R$_B$ = 3.28, |
| O<110> | O | 0.54638 | 0.54638 | 0 | 0.216 | 3 | R$_F$ = 3.87 |

**Table 2.** Model fitting to dielectric data of PCN ceramic sample in various regions.

| Region I | Region II | Region III |
|---|---|---|
| **Cluster glass model** | | **Curie-Weiss law** |
| $\omega_o$ = 2.05 x 10$^8$ Hz | $\omega_o$ = 8 x 10$^8$ Hz | $C$ = 2.6 x 10$^5$ K; $T_{cw}$ = 250 K |
| $T_g$ = 115.7 K | $T_g$ = 199.5 K | **Modified Curie-Weiss law** |
| $zv$ = 3.5 | $zv$ = 2.27 | $\varepsilon_A$ = 4113 ; $T_A$ = 252 K; $\delta_A$ = 90 |

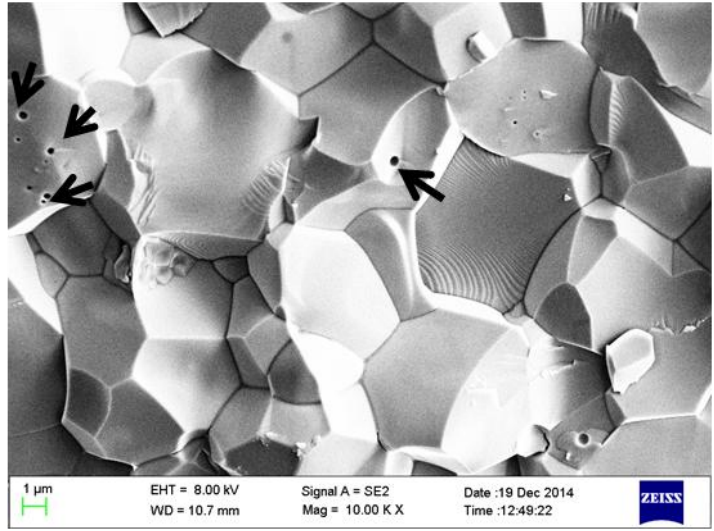

**Fig. 1.** Room temperature SEM micrograph of PCN ceramic sample; entrapment of small spherical pores within the grain is marked by arrow

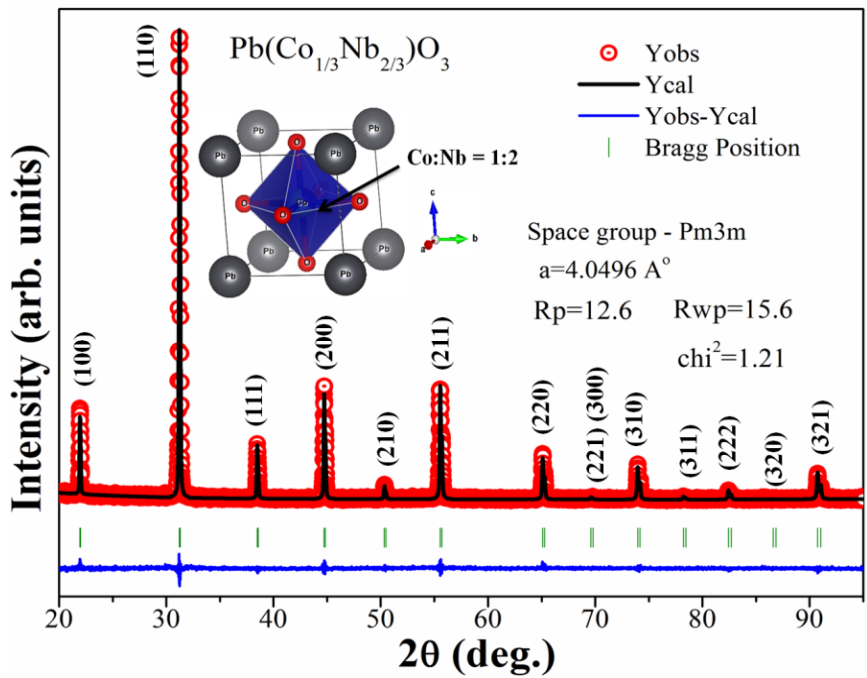

**Fig. 2.** Room temperature fitted XRD pattern of PCN ceramic sample using Pm3m crystal symmetry.

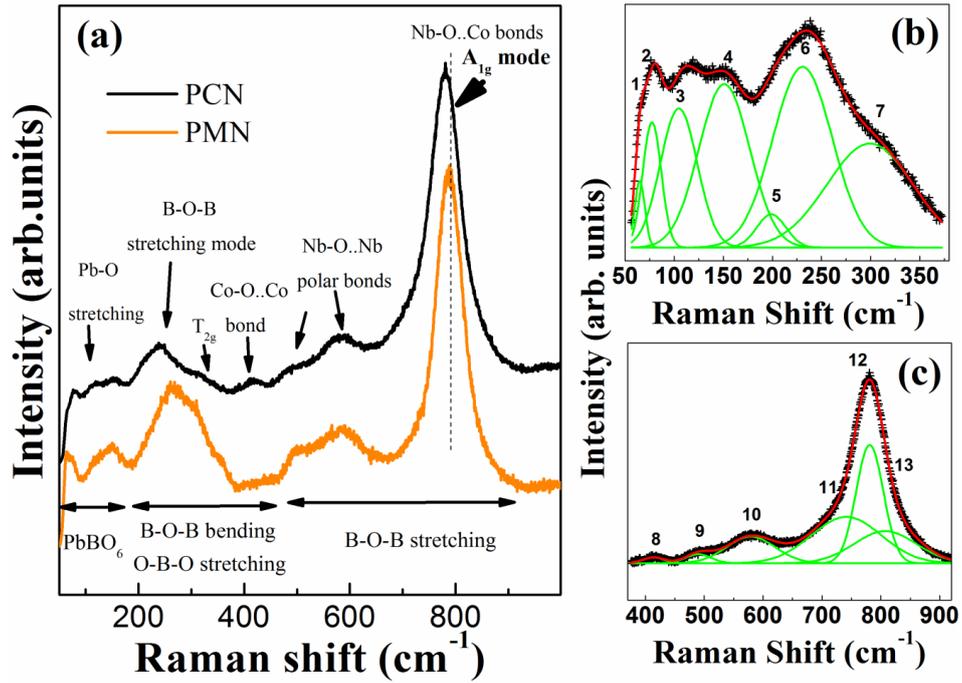

**Fig. 3.** (a) Room temperature Raman spectra of PCN ceramic sample, (b,c) Fitting of Raman spectra using pseudo-voigt peak function.

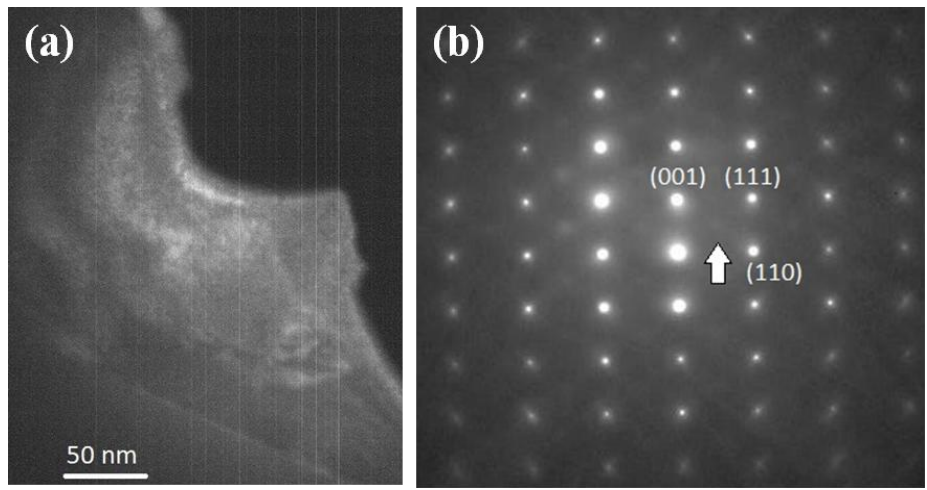

**Fig. 4.** Room temperature (a) bright field image, and (b) SAED pattern along the <110> axis of PCN ceramic sample; strong super-lattice reflections (F-spot) at (½½½) reciprocal positions along <111> are marked by arrow.

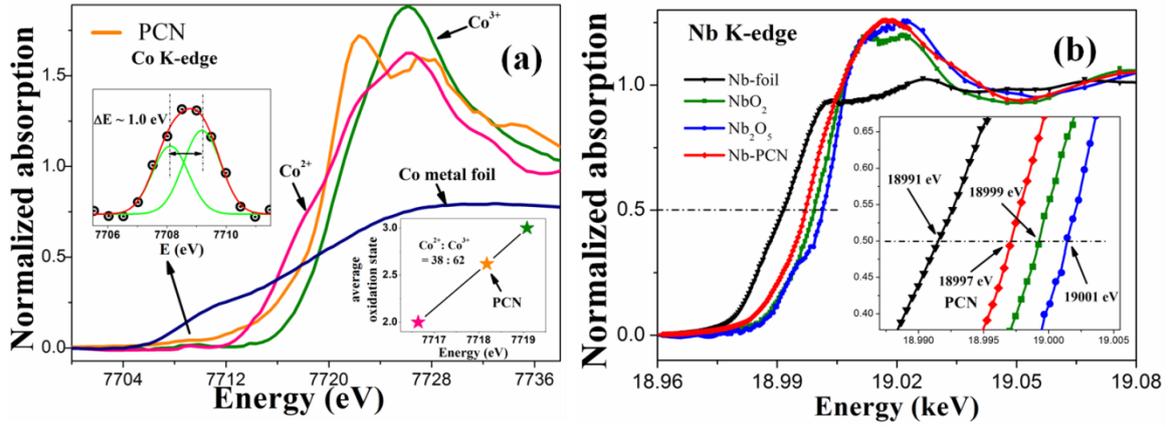

**Fig. 5.** Room temperature (a) Co K-edge and (b) Nb K-edge XANES spectra of PCN ceramic along with their respective standard samples.

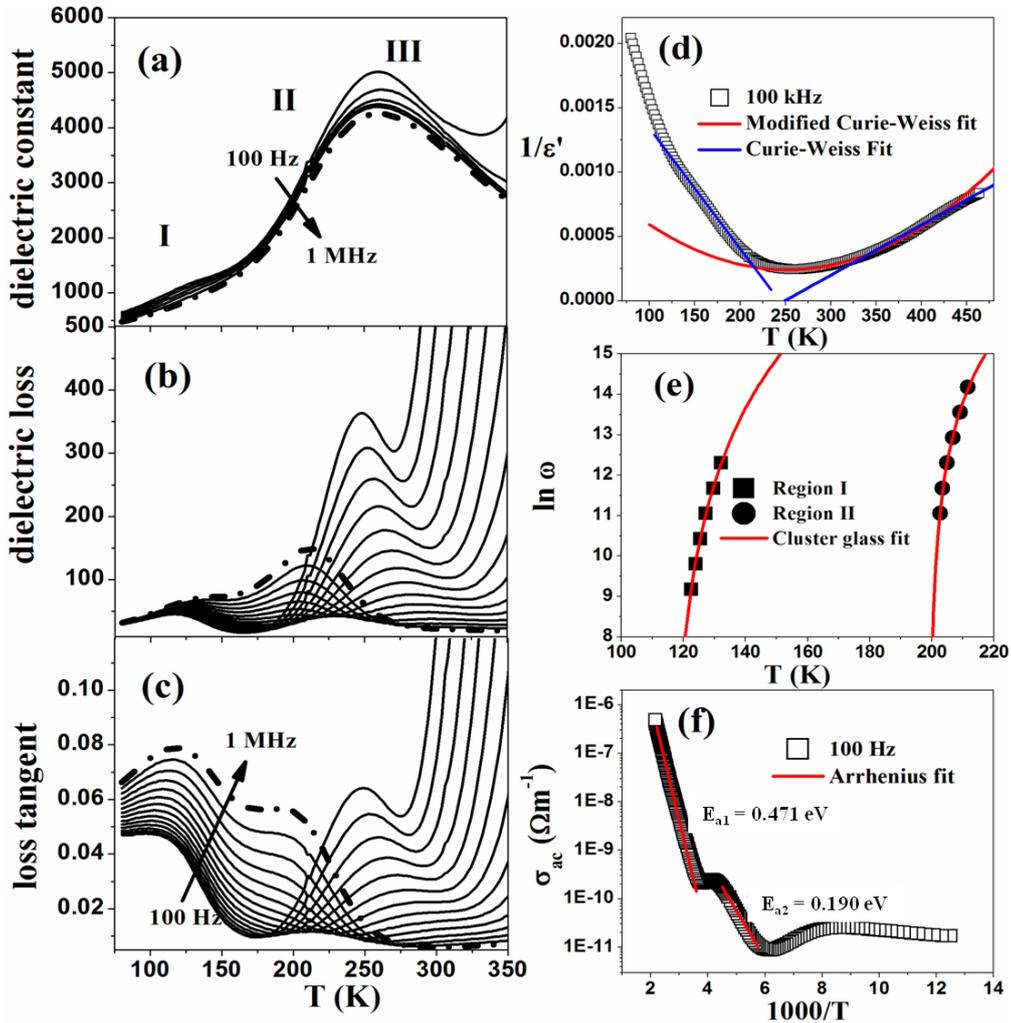

**Fig. 6.** Temperature dependence of (a) real part ($\varepsilon'$), (b) imaginary part ($\varepsilon''$), (c) loss tangent (tan δ) of complex permittivity, (d) Curie-Weiss and Modified Curie-Weiss fit to $1/\varepsilon'$ vs T plot, (e) temperature dependence of relaxation frequency fitted using cluster glass model in region I and II for PCN ceramic sample, and (f) temperature dependence of ac-conductivity of PCN at 100 Hz is fitted by Arrhenius law.

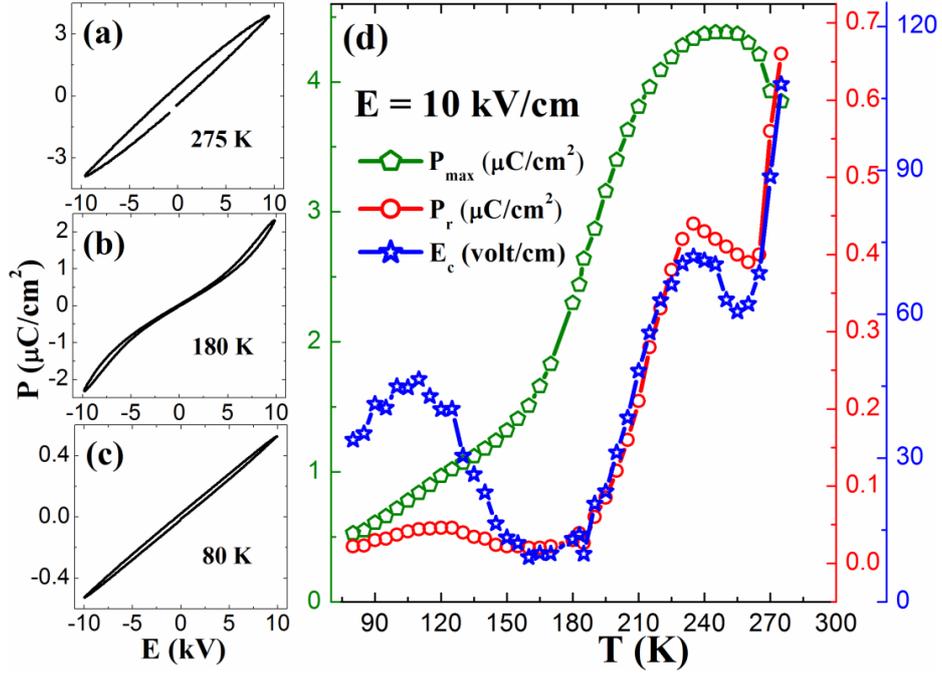

**Fig. 7.** *P-E* hysteresis of PCN ceramic sample at different temperatures, (a) 275 K, (b) 180 K and (c) 80 K, (d) Temperature dependence of $P_{max}$, $P_r$ and $E_c$ of PCN ceramic sample for $E$ = 10 kV/cm applied electric field.

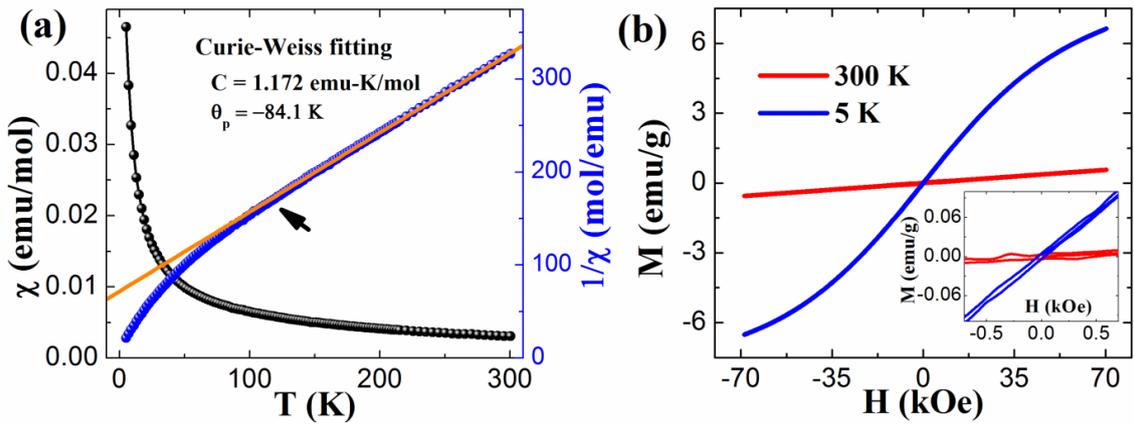

**Fig. 8.** (a) Temperature dependent susceptibility and inverse susceptibility plot of PCN ceramic where high temperature linear region above ~150 K is fitted using Curie-Weiss law, (b) Field dependent magnetization at 300 K and 5 K.

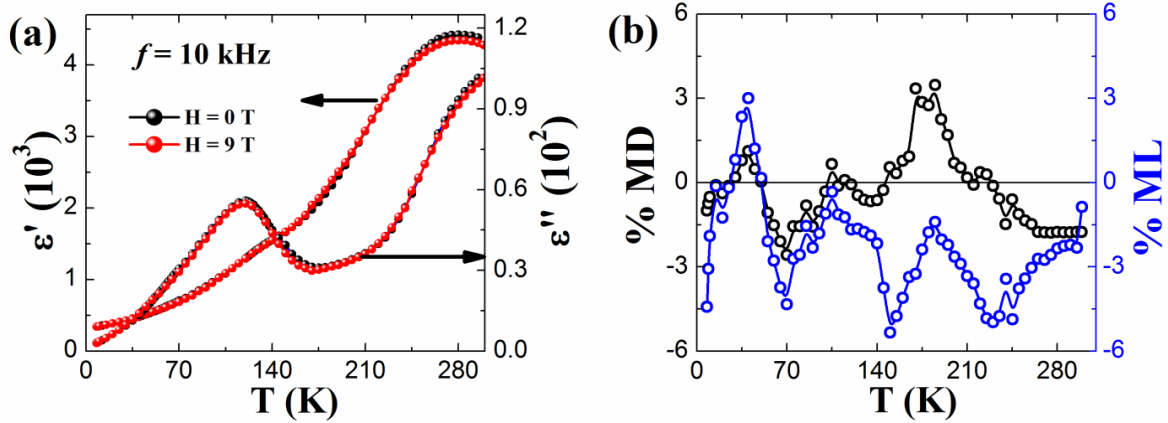

**Fig. 9.** (a) Temperature dependence of real (ε') and imaginary (ε'') parts of complex dielectric permittivity of PCN ceramic sample at H = 0 T and H = 9 T magnetic field at 10 kHz frequency, (b) Temperature dependence of magnetodielectric (%MD) and magnetoloss (%ML) of PCN at 10 kHz